\documentclass[a4paper]{article}
\usepackage{graphicx}
\usepackage{chicago}
\newcommand{\be}{\begin{equation}}
\newcommand{\ee}{\end{equation}}
\newcommand{\bd}{\begin{displaymath}}
\newcommand{\ed}{\end{displaymath}}
\newcommand{\bea}{\begin{eqnarray}}
\newcommand{\eea}{\end{eqnarray}}
\newcommand{\bi}{\begin{description}}
\newcommand{\ei}{\end{description}}
\newcommand{\bq}{\begin{quote}}
\newcommand{\eq}{\end{quote}}

\def\fo{\footnote}

\def\r{\rho}

\def\d{\delta}
\def\D{\Delta}
\def\e{\epsilon}

\def\om{\omega}

\def\l{\lambda}

\def\ph{\varphi}
\begin{document}
\bibliographystyle{chicago}
\twocolumn[
\author{Alexander~Unzicker\\
        Pestalozzi-Gymnasium  M\"unchen\\[0.6ex]
{\small{\bf e-mail:}  alexander.unzicker@lrz.uni-muenchen.de}}

\author{
Alexander~Unzicker\\
         {\small Pestalozzi-Gymnasium M\"unchen, Germany}\\
            {\small aunzicker@lrz.uni-muenchen.de } \\ [4mm]
Karl Fabian\\
         {\small Norwegian Geological Survey Trondheim, Norway}\\
    {\small karl.fabian@ngu.no }}

\date{October 7th, 2006}

\title{A Solar System Test of Mach's Principle with Gravimetric Data}
\maketitle

\begin{abstract}
We present a new test for a possible Mach-Sciama dependence of the
Gravitational constant $G$. According to Ernst Mach (1838-1916),
the gravitational interaction depends on the distribution of
masses in the universe. A corresponding hypothesis of Sciama
(1953) on the gravitational constant, $c^2/G = \sum m_i/r_i$, can
be tested since the elliptic earth orbit should then cause minute
annual variations in $G$. The test is performed  by analyzing the
gravity signals of a network of superconducting gravimeters (SG)
which reach a precision of $10^{-10} m/s^2$. After reducing the
signal by modelling tidal, meteorologic and geophysical effects,
no significant evidence for
 the above dependence is found.

\end{abstract}

\vspace{1.0cm}]

\section{Introduction}

Ernst Mach (1838-1916) suggested that the origin of gravitational
interaction could depend on the presence of all masses in the
universe. After the availability of cosmological data in the
1930s, the approximate coincidence $\frac{c^2}{G}
=\frac{m_u}{r_u}$ (where $m_u, r_u$ refers to mass and radius of
the universe) observed by \citeN{Dir:38a} pushed the development
of that  principle. However, despite its fascinating implications,
very few quantitative statements of Mach's principle have been
developed yet. \citeN{Sci:53} deduced ba a gravitomagnetic analogy
a theory in which a dependence of the gravitational constant on
the mass distribution of the universe, \be \frac{c^2}{G} =\sum
\frac{m_i}{r_i}, \label{sum} \ee is proposed. He estimated the
respective contributions of the solar system, the galaxy, and the
extragalactic matter, found the latter to be greatly dominant and
concluded that the coupling $G$ is `practically constant over the
distances and times available to observation'. Since \be
\frac{M_S}{1 AU} \frac{c^2}{G} \approx 10^{-8}, \ee the elliptic
earth orbit (eccentricity $\e=0.0167$) with an oscillating
distance to the sun leads to a spatiotemporal variation the above
sum (\ref{sum}). If such a variation of $G$ were true, the effect
on local gravity measurements of $g$ would be about  $1.64 \
nm/s^2$, which can be approached by modern superconducting
gravimeters (SG). There are already hints that the hypothesis
(\ref{sum}) is unlikely. For example,
the Hubble expansion 
should cause a drift in $G$ that exceeds the present observational
evidence\fo{See \citeN{Uza} for an overview.} for $\frac{\dot
G}{G} < 10^{-12} yr^{-1}$. However, all these tests rely on
cosmological models, whereas the test described here uses just
celestial mechanics data. Moreover, it is sensitive to a annual
oscillation of $G$ on earth and therefore to a spatial variation
of $G$ rather to a drift, which is usually predicted by the Hubble
expansion.

Though superconducting gravimeters (SG) have an extraordinary
precision, unfortunately a variety of effects practically masks
the detection of a possible signal of the described type. SG time
series have been analyzed recently (\citeNP{Ama:04};
\citeNP{Kro:06}; \citeNP{Boy:06}; \citeNP{Har:06b}), reducing the
raw data with an appropriate modelling of known effects. Due to
difficulties, the results regarding the long-term signal of the
reduced data (gravity residuals) are not yet coherent.

The dominant signal of the direct solar and lunar tides can be eliminated relatively
easily.
The tidal attraction creates then a deformation of the earth which still can be modelled,
though elastic properties of the earth enter here. Considerably more difficult is the calculation
of the deformation of the earth crust due to the tidal shift of oceanic water masses (ocean loading).

Besides these phenomena which occur at well-known frequencies,
meteorologic effects influence the gravity signal. Obviously, air
masses above the SG reduce surface gravity, thus an
anticorrelation with pressure is observed. Again, besides the
direct effect, the
air mass loading creates 
a deformation of the crust which is relevant. Independently of the pressure,
cold air masses, being closer to the SG, have a slightly stronger effect on the signal; thus
temperature data are needed as well. Groundwater level variations show a significant influence
on the data, too.
Very interesting is the extraction of the polar motion signal. Since $g$ varies with latitude $\ph$, a shift of
the rotation axis of the earth effectively changes $\ph$ and therefore the locally measured
$g$.

\section{Data analysis and reduction by modelling}
\subsection{General method}
Our data analysis consists in three steps. After eliminating errors and offsets, in a second step
we try to model all physical effects that influence the gravity signal by avoiding any fitting
parameters, trying to avoid misinterpretation of the remaining residual. In a third step, physical
parameters like the gravimetric factor for polar motion are introduced in a
 least-square-fitting routine and
spectral filtering is performed where appropriate.

\subsection{Raw data preparation}
The raw gravity data are public available at
the web site of the global geodynamics project (GGP)
{\em ggp.gfz-potsdam.de \/} (ask for a guest account).
 Whenever possible, we used the so-called `corrected minute data'
{\em *22.ggp\/} instead of the `minute data'. According to the
description, gaps due to instrument maintenance are filled with
synthetic signal and offsets are adjusted. There are remaining
gaps and offsets however; we corrected the latter ones by
estimating the `jumps' after data reduction described below, then
redoing the automatized analysis. Since our interest was the
long-term-signal, the minute data were filtered and averaged in a
first step to 10 mins. Gaps and corrupted signal was labelled in
order to take it out from the final analysis. Pressure data were
prepared accordingly; in the few cases where absolute pressure
information was missing, it was estimated from mean barometric
pressure of the station height.\fo{For linear regression analysis
done by many groups, the absolute pressure value is indeed
obsolete.}

\subsection{Tidal effects}
Most of the tidal analysis is nowadays done by software
packages (e.g. \citeNP{Wen:96}) that determine factors and
phases for a variety of spectral components. They are
based on the tidal potential computed by \citeN{Har:95}. To separate different physical mechanisms,
we found it interesting instead to use just the public available data of planetary motion.

\paragraph{Direct attraction.}
NASA provides on their `HORIZONS'- website\fo{{\em
http://ssd.jpl.nasa.gov/?horizons \/}} the most precise ephemeris
data, available, DE 405, which is continuously updated with recent
space mission results. Since the other planets are negligible for
our purposes, we used hourly ephemeris data of moon and sun in the
geocentric\fo{Otherwise, one had to deal with earth precession
etc.} coordinate system of equinox 2000, and did a 3D-spline
interpolation to 10 mins. Since the position of the SG station can
be computed from the hour angle and the station height and
latitude (using a ellipsoidal earth model), the vectorial
difference of the gravitational force acting on the earth
barycenter and on the station can be computed easily. Since this
difference is in general not perpendicular to the earth surface,
scalar multiplication with the station normal vector $\vec n$
antiparallel to $\vec g$ is performed. Since SG instruments
automatically adjust themselves into the direction of $\vec g$, a
correction for the centrifugal force is added to the geometric
component of $\vec n$ computed from latitude. For illustration,
fig.~(\ref{solar}) shows the solar component of this direct tide
for one year. The seasonal effect of azimuthal height and the
ellipticity of the earth orbit (slightly changing top line) is
nicely reflected.

\begin{figure}[h]
\caption{Direct solar tide effect for the Moxa station, 2001.}
\label{solar}
\end{figure}

Accordingly, fig.~(\ref{tides_m}) shows the solar and lunar component and its sum for one month (June 2001)
computed for the Moxa station.

\begin{figure}[h]
\caption{Direct solar (green) and lunar (red) tides and their sum (blue) for June 2001 in Moxa.}
\label{tides_m}
\end{figure}

Comparison to the gravity signal from the GGP site,  fig.~(\ref{dirtide_m}), shows that
the direct tidal effect is by far the dominant part of the SG signal.
exemplarily depicted for the Moxa station, June 2001.

\begin{figure}[h]
\caption{Gravity signal (red), direct tide computed from ephemeris data (blue), and the
difference (green) for June 2001 of the Moxa station.}
\label{dirtide_m}
\end{figure}

\paragraph{Earth deformation tides.}
Besides the direct effect, the earth as a deformable body responds
elastically to the direct tide. Instead of determining the a
priori unknown elastic constants of the earth by fitting, as a
first approximation we calculate the deformation by analogy to the
well-known deformation the earth undergoes as a reaction to its
rotation. Assuming linear elasticity\fo{which is justified since
we have geometrically small deviations from the sphere shape}, a
stretch in the equatorial plane induced by the centrifugal force
is equivalent to a squeeze along the pole axis with the doubled
force.\fo{The stretching force acts in two dimensions, the
squeezing force in one.} Instead of squeezing, a gravitation
celestial body stretches the earth along the visual axis to the
body, causing thus a prolate deformation instead of the oblate one
coming from the earth rotation. To compute the gravity effect, we
can simply use the international gravity formula for latitude $\l$
\fo{Numerical values from IUUG.} \be g(\l):=g_0\frac{(1+k_0
\sin(\l)^2)}{\sqrt{1-e_2 \sin(\l)^2}}; \ee with
$g_0=9.7803267714$, $k_0=0.00193185138639$ and
$e_2=0.00669437999013$, and subtract the centrifugal term \be
g_{k}(\l)= g(\l)-a_z(\l). \ee If we take the spherical average of
$g_{k}(\l)$, \be g_{0} = \frac{1}{2 \pi} \int_0^{\frac{\pi}{2}}
\cos(\l) g_{k}(\l) d \l \ee $\bar g_k(\l) := g_{k}(\l)-g_{0}$
describes the gravity due to the oblate deformation only.
If $f_t$ is the maximal tidal force performed 
by the celestial body along the visual axis, then the gravity
effect due to the corresponding prolate deformation computes as
\be \frac{f_t}{2 a_z(0)} \bar g_k(\l), \ee whereby $\l$ now stands
for the deviation angle from the visible axis. $\l$ can again be
computed for the SG station at any time. As as
fig.~(\ref{devtide_m}) shows, the deformation tide is again the
dominant part of the signal remaining in fig.~(\ref{dirtide_m}).
However, the simplified elasticity approach does not take into
account that the centrifugal force, acting for billions of years,
creates much more elastoplastic deformation than the tidal forces
which act for hours only. The above calculation exaggerates
therefore the true deformation by about $15 \%$. Taking this into
account and implementing the well-known phase shift of $3 \deg$ of
the deformation tide, a slightly better reduction of the signal is
obtained, as fig.~(\ref{devphase_m}) shows.

\begin{figure}[h]
\caption{Residual of fig.(\ref{dirtide_m}) (red), deformation tide
computed from prolate deformation (blue), and the remaining
difference (green) for June 2001 of the Moxa station.}
\label{devtide_m}
\end{figure}

\begin{figure}[h]
\caption{With a phase lag of $3 \deg$ and a estimated reduced value 0.85
for the deformation tide, the residual of fig.(\ref{devtide_m}) (red) is
slightly improved (blue).}
\label{devphase_m}
\end{figure}

\paragraph{Residual tide signal.} Despite the improvement in fig.~(\ref{devphase_m}) there
is a clear residual signal at tidal frequencies. The main reason
for it is the tidal shift of oceanic water masses which is not in
phase with the deformation tide and which is greatly influenced by
the distribution of land masses (ocean loading). Naturally,
stations near the coast are more affected. Typical amplitudes
range from $10$ to $100$ $\frac{nm}{s^2}$ There are several models
that calculate tidal frequency components for a given
location.\fo{Online calculation provided by H.-G. Scherneck,
http://www.oso.chalmers.se/~loading/.}
 Since our focus of interest are long-term-effects rather than evaluating
ocean loading models, we preferred to eliminate the remaining tidal signal by spectral filtering.

\subsection{Meteorologic effects}

\paragraph{Pressure reduction.}
Precise surface pressure data for the station location are
provided by the GGP web site (same *.ggp file). To give a first
approximate correction, we calculated the attraction of an
cylindrical air mass of radius 300 km (typical size of atmospheric
disturbances) and 11 km height, using an isothermal barometric
pressure formula with an estimated temperature for the station and
taking the station pressure as overall surface pressure. A
correction for the earth curvature was applied, too. The result is
a clear improvement of the residual signal, see
fig.~(\ref{pressk_m}). Amplitudes for the pressure effect can
reach $50 \frac{nm}{s^2}$.

\begin{figure}[h]
\caption{Residual signal from fig.(\ref{devphase_m}) (red), pressure correction (blue)
and the remaining difference. The residual still contains tidal
frequencies due to ocean loading.}
\label{pressk_m}
\end{figure}

\paragraph{Temperature effect.}

A cold air mass, being closer to the SG, performs 
a greater gravitational attraction than a warm one, even if there
is the same surface pressure. This is much more difficult to
model, and even if temperature recordings of the stations were
desirable, they could not be useful to the same extent, since
temperature shows a greater local variability than pressure.
Therefore, weather data for the region surrounding the SG station
are needed. We used the freely available NCEP Reanalysis
data\fo{The *.nc files were converted into ASCII-readable files by
the freely available programm ncdump.exe.}
 at the web site
{\em http://www.cdc.noaa.gov/ \/}, which provides on a $2.5
\deg$-grid four times daily humidity, temperature and geopotential
height data for 17 pressure levels. For simplicity, we chose a $20
\deg$ x $20 \deg$-grid centered around the SG station and
considered all 17 pressure levels, since for appropriate modelling
3-dimensional data are needed \cite{Neu:04}. Thus even for the
stations closest to the poles a distance of about 500 km to the
station is covered. To summarize, for every station location there
are $9$x$9$x$17=1377$ data points. Every point is at the center of a cuboid 
which is assumed to have homogeneous density. To save computation
time, the gravitational effect of every cuboid is first integrated
analytically for every station, and the resulting weights are than
multiplied with the time-dependent density. Since density data are
not provided, the geopotential height for the pressure level,
$h(p)$, is by vertical interpolation transformed into a function
$p(h)$. Numerical differentiation yields with $\r =\frac{1}{g}
\frac{dp}{dz}$ the density $\r(h)$. This procedure turned out to
be much more efficient and precise than using temperature data and
the general equation of state $p V = N K T$ and $\r = \frac{p}{k T
m}$, where $k$ is the Boltzmann constant and $m$ the molecular
mass. Once the gravitational effect of the surrounding air masses
is computed as described above, the 6h-time series can be
interpolated to 10 mins. Two problems however arise. The station
pressure data is not only available more frequently, but also much
more precise. Thus the loss in accuracy due to the inevitably less
precise NCEP data could spoil the better physical modelling. We
computed therefore an artificial reference pressure $p_{Ncep}$
from the layer model above the station, and set the gravity
correction $g_{Ncep}$ in relation to the true station pressure
$p_{Station}$, thus yielding the final correction \be g_{final}=
g_{Ncep} \frac{p_{Station}}{p_{Ncep}}. \ee Due to temporal
interpolation there is still some high frequency noise introduced
by this procedure, but the gain in accuracy for the long-term
seasonal signal is considerably, as  fig.~(\ref{s111_14c})
exemplarily shows for the Moxa station. During winter season, the
upward air attraction is higher and therefore the signal lower.

\begin{figure}[h]
\caption{Difference between 3D atmospheric density calculation and
pure station pressure correction for the SG in Moxa (Ger) for the
year 2002. Summer and Winter are clearly distinguishable.}
\label{m112_12c}
\end{figure}

The temperature effect can reach as much as $10 \frac{nm}{s^2}$, with an average annual
amplitude up to $5 \frac{nm}{s^2}$.

\paragraph{Air mass loading.}
The pressure distribution on earth creates a deformation of the
crust that again manifests as a gravity change. A method for the
calculation with Greens functions is presented in \citeN{Neu:04}.
Since our global surface pressure data is much less precise than
the station pressure, we restrict here to a linear regression to
the station pressure, since the loading effect is strongly
correlated to it.
Air mass loading can cause effects up to $10 \frac{nm}{s^2}$. 

\paragraph{Hydrology.}
Ground water level variations can have considerable influence on the SG signal,
although a correlation as well as a anticorrelation may occur, depending on
the station location being above (e.g. Sutherland) or below (e.g. Moxa) the
surrounding terrain. However, groundwater and rainfall data are provided by some
stations only. After correcting gaps due to malfunction, we determined the
correlation by least-square fitting, since detailed geological information
about rock porosity etc. is usually missing.
Hydrologic effects are highly station-dependent and may reach $50 \frac{nm}{s^2}$
in extreme cases.

\subsection{Geodynamic signals.}

The rotation axis of the earth does not always point through the same point
but shows an oscillation with an amplitude of some tens of meters (polar motion).
Such a shift effectively changes the latitude of the SG station, and consequently,
one expects a change in the centrifugal force that has usually a component
perpendicular to the station surface. This expected change is amplified by the
so called gravimetric factor $\d$ (\citeNP{Loy:99}; \citeNP{Duc:06}),
since a change in latitude causes a variation of $g$ independently from
the centrifugal force (see the above discussion on the international gravity
formula).

\begin{figure}[h]
\caption{Example for polar motion signal (dark green) and residual
signal after fitting for Sutherland, 2001-2004.}
\label{s111_14_red}
\end{figure}

Due to VLBI techniques, very precise data are provided by the
International Earth Orientation Service (www.IERS.org). It is easy
to compute he effective change in latitude for the SG station and
the expected form of the signal. We then determined the
gravimetric $\D$  factor by fitting. It is possible that $\d$
varies slightly with position. We do not know however any physical
reason why $\d$ should vary considerably between the Chandler
frequency (432 d) and the annual one (365.25 d). Fitting is done
therefore with respect to the complete signal, not to separate
frequencies.

\paragraph{Length-of-day variations (LOD).}

\begin{figure}[h]
\caption{Predicted gravity changes due to length-of-day-variations
(red, 10 times exaggerated), and polar motion signal, for Moxa,
2001.} \label{lodch}
\end{figure}

Variations of the length of the day and therefore of the rotational frequency $\om$
cause variations of the centrifugal force. Though this is in the range below
$2 \frac{nm}{s^2}$, we included the LOD data in our analysis.

\subsection{Drift.}
All SG instruments show a drift that is not fully understand yet,
thus absolute gravity measurements are not possible. The drift has
to be eliminated by fitting a linear function in time.

\begin{figure}[h]
\caption{Signal with and without polar motion before removing the drift.}
\label{chandler_s}
\end{figure}

\section{Results}

\subsection{Polar motion.}

The polar motion signal is obtained after applying tide, pressure
and temperature corrections and drift removal. For the plot
fig.~(\ref{chall}), tidal frequencies are filtered. The
gravimetric factors $\d$ obtained by least-square fitting are
shown in table~2. $\d =1.$ would correspond to the centrifugal
force change only, but the elastic response has to be taken into
account, too. Our result agrees with \shortciteN{Kro:06} for Moxa,
but $\d$ varies significantly with location. Contemporary fitting
with pressure and temperature corrections does not change very
much. Additional fitting to the groundwater level signal yields
still lower values for $\d$, $1.09$ and $1.12$ for Moxa and
$1.02$ and $1.03$ for Sutherland (for Bolder and Canberra, no
hydrologic data are available).

\vspace{1.0cm}

\begin{tabular}{|lccr|}
\hline
Station & temperature corr. & pressure only & \\
\hline
Mo & 1.10 & 1.13&\\
Su & 1.06 & 1.07 &\\
Bo & 1.40 & 1.41 &\\
Cb & 1.21& 1.22 &\\
\hline
\end{tabular}

\vspace{0.5cm}

Table~1.\\
Gravimetric factor $\d$ for different stations obtained by fitting
the polar motion signal, together with either the temperature
correction (left) or the pressure correction only (right).

\begin{figure}[h]
\caption{Polar motion signal after removing tides, pressure and temperature corrections,
for the stations Moxa, Sutherland, Bolder, Canberra. The respective gravimetric factors
obtained from fitting are shown in Table~1, left column.}
\label{chall}
\end{figure}

\subsection{Temperature corrections.}

The 3D- density correction shows a clear annual signal resulting from different seasonal
atmospheric densities. In Winter, the cold atmosphere over Moxa and Bolder has a stronger
effect, whereas for the stations on the southern hemisphere the opposite effect is
observed.

\begin{figure}[h]
\caption{Difference between the simple pressure correction and the density correction in a 3D-model
obtained from NCEP climate data,
for the stations Moxa, Sutherland, Bolder, Canberra.}
\label{teall}
\end{figure}

Since the temperature correction as described in section~2.4 introduces some high-frequency
noise, the fitting quality is not improved (see Table~2).

\vspace{1.0cm}
\begin{tabular}{|lccccr|}
\hline
Station & T & RMS & P & RMS& \\
\hline
Mo & 0.76 & 11.70& 0.89 & 11.66&\\
Su & 0.81 & 10.36 & 1.00 & 10.32&\\
Bo & 0.75 & 16.12 & 0.93 & 16.13&\\
Cb & 0.70 & 26.99  & 0.84& 27.0 &\\
\hline
\end{tabular}\\

\vspace{0.5cm}

Table~2.\\
`TC' stands for temperature correction, and `PR' for pressure only. Due to insufficient
temporal resolution of climate data, the high-frequency noise spoils the long-term
improvement, as least-square deviations from the signal are concerned.

\subsection{Machian signal}

Since the possible Machian dependence has the same structure as it
might be expected from other seasonal signals related to climate
effects, there is considerable danger for misinterpretation. In
particular, if such an influence is not modelled, the
temperature-dependent atmospheric density may mimic a Machian
signal {\em on the northern hemisphere\/}. Indeed, the respective
stations show a high value in the middle column of table.~3 where
temperature
 is not modelled. It is expected therefore
that the temperature correction reduces the Machian factor, as it
 is observed for Moxa and Bolder.
There is however a positive correlation with the Machian signal
for Sutherland and Canberra, too, which cannot be a
misinterpretation of climate seasons, since these
 effects should be reversed.

Since polar motion occurs at similar frequencies as well, a
badly estimated $\d$ could be
the reason for a fake Machian signal. With a $\d=1.16 $
fixed we find a positive correlation
to the Machian signal as well, though with considerable variation.
\vspace{0.5cm}

\begin{tabular}{|lcccr|}
\hline
Station & TC & PR & PF&\\
\hline
Mo & 2.49 & 3.43& 1.13 &\\
Su & 1.74 & 1.25 & 2.12&\\
Bo & 3.16 & 3.46 & 4.25&\\
Cb & 2.37 & 2.65 & 2.46&\\
\hline
\end{tabular}\\

\vspace{0.5cm}

Table~3.\\
Fitting factor for the Mach-Sciama signal for the four stations under consideration.
A factor $1.$ would indicate an annual signal of amplitude $1.64 \frac{nm}{s^2}$ with a
minimum at the perihelion (Dec).
`TC' stands for temperature correction, `PR' for pressure only, and `PF' for polar motion fixed.
In the left and middle column, fitting is done contemporarily to atmospheric correction signal, the polar motion signal
and the Mach-Sciama-signal. In the right column, polar motion is kept fixed to $\d=1.16$.\\

\begin{figure}[ht]
\caption{Residual signal after removing tides, pressure and temperature corrections,
and polar motion, low-pass-filtered,
for the stations Moxa, Sutherland, Bolder, Canberra}
\label{machall}
\end{figure}

To put in evidence the difficulties in eliminating unwanted
signals, fig.~(\ref{machall}) shows the residual after applying
corrections for tides, polar motion, pressure and temperature.
Additionally, since we are interested in the long-term components
of the signal, all frequencies $1/150 d$ were cut off. The
low-frequency noise due to unknown effects like station
maintenance, instrument failures, insufficiently corrected gaps,
drifts and offsets is still above the amplitude of the Machian
signal, $1.64 \frac{nm}{s^2}$. Therefore we cannot attribute
significant evidence to the fitting coefficients found above.

\section{Discussion and Outlook}

The large variation of the gravimetric factors for different
locations has to be investigated further. While we do not see any
reason why $\d$ should vary with frequency, local changes must be
considered. Since polar motion is an effective change in latitude,
gravity gradients having its origin in the mantle and core may
cause such a variation.

We have shown how to analyze a possible Mach-Sciama-signal, in
particular the comparison of the two hemispheres and climate
modelling turns out to be important. Though we find a little
signal with the correct sign for all stations, the overall noise
is yet too large for attributing significance to it. A very
unlikely hypothesis like the Mach-Sciama-dependence of $G$ with
its consequences for Newton's law needs much stronger evidence.
Rather than refining the statistical analysis, we propose future
efforts in improving the physical models. The found polar motion
anomaly should be investigated first. To exclude climate effects,
(possibly dry)  SG stations on the southern hemisphere are
desirable.

\paragraph{Acknowledgement.}
Gravity and pressure data are provided by the GGP network
http://ggp.gfz-potsdam.de. NCEP climate reanalysis data are
provided by the NOAA/OAR/ESRL PSD, Boulder, Colorado, USA, from
their web site at http://www.cdc.noaa.gov/. Planetary ephemeris
data is provided by NASA from the web site
http://ssd.jpl.nasa.gov/?horizons. Earth orientation and LOD data
are provided by http://www.iers.org. We are grateful to Corinna
Kroner and to Thomas Kl\"ugel for helpful hints and suggestions.

\end{document}